\documentstyle[12pt]{article}

\large
\begin{document}
\setlength{\baselineskip}{21pt}

\vskip 1.0cm
\begin{center}
{\Large No $Z_N$ - bubbles in hot Yang-Mills theory}
\footnote{Invited talk presented
on the International Conference on High Energy Physics,
 Glasgow, July 94.} \\
\vskip 1cm
A.V. Smilga\\
\vskip 0.4cm
ITEP, B.Cheremushkinskaya 25, Moscow 117259, Russia
\vskip 1cm
\end{center}
\centerline{\bf Abstract}

Pure Yang-Mills theory at high temperature is considered. We show
that no distinct $Z_N$- phases separated by domain walls do exist in the
physical Minkowski space. That means the absense of the spontaneous breaking
of $Z_N$- symmetry in the physical meaning of this word.
\vskip 1cm

\section{Introduction.}
It was shown some time ago that the pure YM theory undergoes a phase transition
at some temperature $T_c \sim \Lambda_{QCD}$ \cite{Pol,Sus}.
This phase transition exhibits
itself in a radical change of the behaviour of the correlator
 \begin{equation}
 \label{corr}
C({\bf x}) = <P({\bf x}) P^*(0)>_T
 \end{equation}
where $P({\bf x})$ is the Polyakov line
 \begin{equation}
 \label{p}
P({\bf x}) = \frac 1{N_c} {\rm Tr} \{ \exp[ig\beta \hat{A}_0({\bf x}) ] \}
 \end{equation}
(we choose the gauge where $\hat{A}_0$ is time-independent; $\beta = 1/T$).
Physically, this gauge transition corresponds to deconfinement: at low $T$,
the interaction part of free energy of a test heavy quark-antiquark pair at
distance $R$ grows linearly with $R$ whereas, for high $T$ , it tends to zero
at large distances.

There were scores of papers published since 1978 where it was explicitly or
implicitly assumed that one can use the cluster decomposition for the
correlator
(\ref{corr}) at large $T$ and attribute the meaning to the temperature average
$<P>_T$. Under this assumption, the phase of this average can acquire $N_c$
different values:
$<~P>_T$ $= C\exp\{2\pi ik/N_c\}, \ \ k = 0, \ldots ,N_c-1$ which would
correspond
 to $N_c$ distinct physical phases and to the
spontaneous breaking of the discrete $Z_N$ - symmetry. In recent \cite{pis},
the surface energy density of the domain walls separating these phases has been
evaluated.

We show, however, that the standard interpretation is wrong. In particular:
\begin{enumerate}
 \item Only the correlator (\ref{corr}) has the physical meaning. The
phase
of the expectation value $<P>_T$ is not a physically measurable quantity.
There is only {\em one} physical phase in the hot YM system.
 \item The "walls" found in \cite{pis} should not be interpreted as physical
objects living in Minkowski space but rather as Euclidean field configurations,
kind of "planar instantons" appearing due to nontrivial $\pi_1[{\rm {\cal G}}]
= Z_N$ where {\cal G}$= SU(N)/Z_N$ is the true gauge symmetry group of the
{\em pure} YM system.
  \item The whole bunch of arguments which is usually applied to nonabelian
theories can be transferred with a little change to hot QED. The latter also
involves planar instantons appearing due to nontrivial $\pi_1[U(1)] = Z$.
These instantons should {\em not}, however, be interpreted as Minkowski space
walls.
\end{enumerate}

It is impossible to present an adequate discussion of this issue in this short
note. The reader is referred to \cite{ja} where such a discussion is given.
We can only briefly mention here some crucial points of our reasoning.

\section{Continuum Theory.}
A preliminary remark is that the situation when the symmetry is broken at high
temperatures and restores at low temperatures is very strange and unusual.
The opposite is much more common in physics. We are aware of only one model
example where spontaneous symmetry breaking survives and can even be induced
at high temperatures \cite{Moh}. But the mechanism of this breaking is
completely different from what could possibly occur in the pure Yang-Mills
theory.

Speaking of the latter, we note first that
there is no much sense to speak about the spontaneous breaking of
$Z_N$ - symmetry because such a symmetry is just not there in the theory.
As was already mentioned, the true gauge group of pure YM theory is
$SU(N)/Z_N$ rather than $SU(N)$. This is so because the gluon fields belong
to the adjoint colour representation and are not transformed at all under the
action of the elements of the center $Z_N$ of the gauge group $SU(N)$.

$<P>_T$ as such is not physical because it corresponds to introducing a
single fundamental source in the system: $<P>_T\  = \exp\{-\beta F_T\}$
where
$F_T$ is the free energy of a single static fundamental source \cite{Larry}
. But one {\em can}not put a single fundamental source
in a finite spatial box with
periodic boundary conditions \cite{Hift}. This is due to the Gauss law
constraint: the total colour charge of the system "source + gluons in the
heat bath" should be zero, and adjoint gluons cannot screen the fundamental
source. This observation resolves the troubling paradox: complex $<P>_T$
would mean the complex free energy $F_T$ which is meaningless.

The "states" with different $<P>_T$ could be associated with different minima
of the effective potential \cite{Weiss}
 \begin{equation}
 \label{pot}
V_T^{eff}(A_0^3) = \frac{\pi^2T^4}{12}\left\{ 1 - \left[ \left(
\frac{gA_0^3}{\pi T} \right)_{mod.2} - 1 \right]^2 \right\}^2
 \end{equation}
For simplicity, we restrict ourselves here and in the following
with the $SU(2)$ case.

This potential
is periodic in $A_0^3$. The minima at $A_0^3 = 4\pi nT/g$ correspond to $P=1$
while the minima at $A_0^3 = 2\pi (2n+1)T/g$ correspond to $P = -1$. There
{\em are} also planar (independent of y and z) configurations which interpolate
between $A_0^3 = 0$ at $x = -\infty$ and $A_0^3 = 2\pi T/g$ at $x = \infty$.
These configurations contribute to Euclidean path integral and are
topologically
non-equivalent to the trivial configuration $A_0^3 = 0$
(Note that the configuration interpolating
between $A_0^3 = 0$ and $A_0^3 = 4\pi T/g$ {\it is} topologically equivalent
to the trivial one. Such a configuration corresponds
to the equator on $S^3 \equiv
SU(2)$ which can be  easily slipped off. A topologically nontrivial
configuration corresponds to a meridian going from the north pole of the sphere
to its south pole and presents a noncontractible loop on $SU(2)/Z_2$ ).
Actually, such configurations were known for a long time by the nickname of
't Hooft fluxes \cite{Hooft}.

 Minimizing the surface
action density in a nontrivial topological class, we arrive at the
configuration
which is rather narrow (its width is of order $(gT)^{-1}$) and has the action
density
  \begin{equation}
  \label{act}
\sigma^{su(2)} = \frac{4\pi^2 T^2}{3\sqrt{3} g} + CgT^2
  \end{equation}
(the constant $C$ cannot be determined analytically in contrast to the claim
of \cite{pis} due to infrared singularities characteristic for thermal gauge
theories \cite{Linde}). These topologically nontrivial Euclidean configurations
are quite analogous to instantons. Only here they are delocalized in two
tranverse directions and thereby the relevant topology is determined by
$\pi_1[{\rm {\cal G}}]$ rather than $\pi_3[{\rm {\cal G}}]$ as
for usual localized instantons. But, by the same token as the instantons
cannot be interpreted as real objects in the Minkowski space even if they
are static (and, at high $T$, the instantons with the size $\rho \gg T^{-1}$
become static), these planar configurations cannot be interpreted as real
Minkowski space domain walls.

I want to elucidate here the analogy between nonabelian and abelian theories.
The effective potential for standard QED at high temperature has essentially
the same form as (\ref{pot}):
   \begin{equation}
 \label{potab}
V_T^{eff}(A_0) = -\frac{\pi^2T^4}{12}\left\{ 1 - \left[ \left(
\frac{eA_0}{\pi T} + 1 \right)_{mod.2} - 1 \right]^2 \right\}^2
 \end{equation}
It is periodic in $A_0$ and acquires minima at $A_0 = 2\pi nT/e$. Here
different
minima correspond to the same value of the standard Polyakov loop
$P_1( {\bf x}) = \exp\{ie\beta A_0( {\bf x})\}$. One can introduce
, however, the quantity $P_{1/N}( {\bf x}) =
\exp\{ie\beta A_0({\bf x})/N\}$ which corresponds to probing the system
with a fractionally charged heavy source : $e_{{\rm source}} = e/N$. Note that
 a fractional heavy source in a system involving
only the fermions with charge $e$ plays exactly the same role
 as  a fundamental heavy source in the pure YM system involving
only the adjoint colour fields. A single fractional source would distinguish
between different minima of the effective potential. If $N \rightarrow
\infty$ , all minima would be distinguished, and we would get infinitely many
distinct "phases".

But this is wrong. One cannot introduce a {\em single}
fractional source  and measure $<P>_T$ as such due to the Gauss law constraint.
What can be done is to introduce a pair of fractional charges with opposite
signs and measure the correlator $<P_{1/N}({\bf x}) P_{1/N}^*(0)>_T$.
The latter is
a physical quantity but is not sensitive to the phase of $P$. The same concerns
the correlator $<P_{1/N}({\bf x}_1) \ldots P_{1/N}({\bf x}_N)>_T$ which
corresponds to putting $N$ fractional same-sign charges at different spatial
points.

Finally, one can consider the configurations $A_0({\bf x})$ interpolating
between different minima of (\ref{potab}). They are topologically inequivalent
to trivial configurations and also have the meaning of planar instantons
\footnote{In the abelian case, there are infinitely many topological classes:
 $\pi_1[U(1)] = Z$.}.
But not the meaning of the walls separating distinct physical phases.
The profile and the surface action density of these abelian planar instantons
can be found in the same way as it has been done in Ref.\cite{pis} for
the nonabelian case. For configurations interpolating between adjacent
minima, one gets
  \begin{equation}
  \label{sab}
\sigma^{u(1)} = \frac{2\pi^2 (2\sqrt{2} - 1)T^2}{3\sqrt{6}e}
+ CeT^2 \ln(e)
  \end{equation}
where $C$ is a numerical constant which {\it can}
in principle be analytically evaluated.

There is a very fruitful and
instructive analogy with the Schwinger model.  Schwinger model is
the two-dimensional $QED$ with one massless fermion. Consider this theory
at high temperature $T \gg g$ where $g$ is the coupling constant (in
two dimensions it carries the dimension of mass). The effective potential in
the constant $A_0$ background has the form which is very much analogous to
(\ref{pot},\ref{potab}):
  \begin{equation}
  \label{potSM}
V^{\rm eff}(A_0) = \frac{\pi T^2}{2}\left[ \left(1 + \frac {gA_0}{\pi T}
\right)^2 - 1 \right]^2
  \end{equation}
It consists of the segments of  parabola and is periodic in $A_0$
with the period
$2\pi T/g$. Different minima of this potential are not distinguished by a heavy
integerly charged probe but could be distinguished by a source with
fractional charge. Like in four dimensions, there are topologically nontrivial
field configurations which interpolate between different minima. These
configurations are localized (for $d=2$ there are no transverse directions over
which they could extend) and are nothing else as high-$T$ instantons. The
minimum of the effective action in the one-instanton sector is achieved
at the configuration \cite{ja,inst}
 \begin{equation}
  \label{inst}
A_0(x) = \left[ \begin{array}{c} \frac {\pi T}g \exp\left\{
\frac g{\sqrt{\pi}}(x - x_0) \right\}, \ \ \ x \leq x_0 \\
\frac {\pi T}g \left[ 2- \exp\left\{
\frac g{\sqrt{\pi}}(x_0 - x) \right\}\right], \ \ \ x \geq x_0
\end{array}
\right.
 \end{equation}
the instanton (\ref{inst}) is localized at distances $x - x_0 \sim g^{-1}$
and has the action $S_I = \pi^{3/2}T/g$. But, in spite of that it is
time-independent, it is the essentially Euclidean configuration and should
not be interpreted as a "soliton" with the mass $M_{sol.?} = TS_I$ living
in the physical Minkowski space.

\section{Lattice Theory}
The most known and the most often quoted arguments in favour of the standard
conclusion of the spontaneous breaking of $Z_N$-symmetry in hot Yang-Mills
theory come from lattice considerations. Let us discuss anew these arguments
and show that, when the question is posed properly, the answer {\it is}
diferent.

Following Susskind \cite{Sus}, consider the hamiltonian lattice formulation
where the theory is defined on the 3-dimensional spatial lattice and the time
is continuous. In the standard formulation, the dynamic variables present the
unitary matrices $V({\bf r}, {\bf n})$ dwelling on the links of the lattice
(the link is described as the vector starting from the lattice node {\bf r}
with the direction {\bf n}).
The hamiltonian is
 \begin{equation}
 \label{hamlat}
H = \sum_{\rm links} \frac{g^2(E^a)^2}{2a} - \frac 2{ag^2} \sum_{\rm plaq.}
{\rm Tr}\{V_1 V_2 V_3 V_4\}
 \end{equation}
where $a$ is the lattice spacing, $g$ is the coupling constant and $E^a$ have
the meaning of canonical momenta $[E^a({\bf r}, {\bf n}), V({\bf r}, {\bf n})]
= t^a V({\bf r}, {\bf n})$. Not all eigenstates of the hamiltonian
(\ref{hamlat}) are, however, admissible but only those which satisfy the Gauss
law constraint. Its lattice version is
  \begin{equation}
  \label{Gaulat}
G^a({\bf r}) = \sum_{\bf n} E^a({\bf r}, {\bf n}) = 0
  \end{equation}
It is possible to rewrite the partition function of the theory (\ref{hamlat},
\ref{Gaulat}) in terms of the {\it dual variables} $\Omega_{\bf r} \in SU(2)$
which are defined not at links but at the nodes of the lattice.
$\Omega_{\bf r}$ are canonically conjugate to the Gauss law constraints
(\ref{Gaulat}) and have the meaning of the gauge transformation matrices
acting on the dynamic variables $V({\bf r}, {\bf n})$.
In the strong
coupling limit when the temperature is much greater than the ultraviolet
cutoff $\Lambda_{ultr} \sim 1/a$, the problem can be solved analytically.
The effective dual hamiltonian has 2 sharp minima at
$\Omega_{\bf r} = 1$ and $\Omega_{\bf r} = -1$ and this has been
interpreted as the
spontaneous breaking of $Z_2$-symmetry.

Note, however, that the same arguments could be repeated in a much simpler and
the very well known two-dimensional Ising model. Being formulated in terms of
the physical spin variables $\sigma$, the theory exhibits the spontaneous
breaking of $Z_2$-symmetry at low temperatures, and at high $T$ the symmetry
is restored. But the partition function of the Ising model can also be
written in terms of the dual variables $\eta$ defined at the plaquette
centers \cite{Kram}.
Dual variables are ordered at
high rather than at low temperatures. This
obvious paradox is resolved by noting that the dual variables $\eta$ are
not measurable and have no direct physical meaning. The "domain wall"
configurations interpolating between $\eta = 1$ and $\eta = -1$ {\it do}
contribute in the partition function formulated in dual terms. But one cannot
feel these configurations in any physical experiment.

And the same concerns the lattice pure YM theory . There {\it are}
configurations interpolating between different $\Omega{\bf r} \in Z_2$
 and contributing to the partition function, but they do not correspond
to any real-time object and cannot be felt as such in any physical
experiment.

Up to now we discussed the system with a standard lattice hamiltonian
(\ref{hamlat}). Note, however, that one can equally well consider the lattice
theory with the hamiltonian having the same form as (\ref{hamlat})
but involving
not the unitary but the orthogonal matrices $V^{adj}({\bf r}, {\bf n})
 \in SO(3)$. Both lattice
theories should reproduce one and the same continuous Yang-Mills theory
in the limit when the inverse lattice spacing is much greater than all physical
parameters (As far as I understand, there is no unique opinion on this issue
in the lattice community. If, however, lattice hamiltonia involving unitary
and orthogonal matrices would indeed lead to different field theories in the
continuum limit, it would mean that the Yang-Mills field theory is just not
defined until a particular procedure of ultraviolet regularization is
specified.
This assertion seems to me too radical, and I hesitate to adopt it.).

But in the strong coupling limit $T \gg \Lambda_{ultr.}$ the two lattice
theories are completely different. The theory with orthogonal matrices has
the same symmetry properties as the continuum theory , and there is no
$Z_2$-symmetry whatsoever. The effective dual hamiltonian depending on the
gauge transformation matrices $\Omega_{\bf r}^{adj} \in SO(3)$ also has no such
symmetry and there is nothing to be broken.

 Earlier the lattice studies of the deconfinement phase transition have been
performed exclusively with the standard lattice lagrangian involving unitary
matrices. These studies suggest that the deconfinement phase transition
occurs simultaneously with the spontaneous symmetry breaking in the dual
hamiltonian  $H^{eff}(\Omega_{\bf r}^{fund.})$  \cite{lat}
(We repeat that such a breaking is not a physical symmetry breaking because
it does not lead to the appearance of domain walls detectable in experiment.).
In our opinion, however, the additional $Z_2$-symmetry which the
hamiltonian (\ref{hamlat}) enjoys is a nuissance rather than an advantage.
It is a specifically lattice feature which is not there in the continuum
theory. We strongly suggest to people who can do it to perform a numerical
study of the deconfinement phase transition for the theory involving
orthogonal matrices. In that case, no spontaneous $Z_N$ breaking can
occur. Probably, for finite lattice spacing, one would observe kind of
crossover rather than the phase transition. The crossover is expected to
become more and more sharp  as the lattice spacing (measured in physical
units) would become smaller and smaller.

It would be interesting also to try to observe the "walls" (i.e. the planar
Euclidean instantons) for the orthogonal lattice theory. They should
"interpolate" between
$\Omega_{\bf r}^{adj} = 1$ and $\Omega_{\bf r}^{adj} = 1$
along a topologically nontrivial
path. Like any other topological effect, these instantons should become
visible only for a small enough lattice spacing (much smaller than the
characteristic instanton size), and to detect them
 is definitely not an easy task.
But using the orthogonal matrices
 is the only way to separate from lattice artifacts. The only available
numerical study \cite{Kaj} was done for the theory with unitary matrices
and  too close to  the strong coupling regime where these artifacts are
desisive. Thereby, it is not conclusive.

\section{Acknowledgements.}
I benefited a lot from numerous discussions with many people. An incomplete
list includes P. Hasenfratz, C. Korthals-Altes, H. Leutwyler,
M. L\"{u}scher, L. McLerran, F. Niedermayer, L. Susskind, N. Weiss and
U. Wiese . It is a pleasure to thank the organizers of this conference
for their excellent and dedicated job and for kind hospitality.


\begin{thebibliography}{20}
\bibliographystyle{unstr}

\bibitem{Pol} A.M.\ Polyakov, Phys. Lett. {\bf 72B} (1978) 477.
\bibitem{Sus} L.\ Susskind, Phys. Rev. {\bf D20} (1979) 2610.
\bibitem{pis} T.\ Bhattacharya  {\it et al.}, Nucl. Phys. {\bf B383} (1992)
497.
\bibitem{ja} A.V.\ Smilga, Ann. Phys., {\it to be published} .
\bibitem{Moh} R.N.\ Mohapatra and G.\ Senjanovic, Phys. Rev. {\bf D20}
(1979) 3390.
\bibitem{Larry} L.\ McLerran and B.\ Svetitsky, Phys. Rev. {\bf D24}
(1981) 450; \\
J.\ Kuti, J.\ Polonyj and K.\ Szlachanyj, Phys. Lett. {\bf 98B} (1981) 199.
\bibitem{Hift} E.\ Hift and L.\ Polley, Phys. Lett. {\bf 131B}(1983) 412.
\bibitem{Weiss} N.\ Weiss, Phys. Rev. {\bf D24} (1981) 475.
\bibitem{Hooft} G.\ 't Hooft, Nucl. Phys. {\bf B153} (1979) 141;
Acta Physica Austriaca Suppl. {\bf 22} (1980) 53.
\bibitem{Linde} A.\ Linde, Phys. Lett., {\bf 93B} (1980) 327;
  D.J.\ Gross, R.D.\ Pisarski and L.G. Yaffe, Rev. Mod. Phys. {\bf 53}
(1981) 43.
\bibitem{inst} A.V.\ Smilga, Phys. Rev. {\bf D49} (1994) 5480.
\bibitem{Kram} H.A.\ Kramers and G.H.\ Wannier, Phys. Rev. {\bf 60} (1941)
252;\\
L.P.\ Kadanoff and H.\ Ceva, Phys. Rev. {\bf B3} (1971) 3918.
\bibitem{lat} J.\ Fingberg, U.\ Heller and F.\ Karsch, Nucl.Phys.
{\bf B392} (1993) 493 {\it and references therein}.
\bibitem{Kaj} K.\ Kajantie et al., Nucl. Phys. {\bf B357} (1991) 693.
\end{thebibliography}
\end{document}